\documentclass[aps,pre,superscriptaddress,longbibliography]{revtex4-1} 
\usepackage{graphicx,amsmath}
\usepackage{hyperref}
\usepackage{amssymb}
\usepackage{gensymb}
\usepackage{color}
\usepackage{placeins}
\usepackage{latexsym}
\usepackage{tikz}
\usetikzlibrary{arrows,decorations.markings}
\usepackage{pgfplots}

\definecolor{forestgreen}{rgb}{0, 0.49, 0}
\definecolor{royal}{RGB}{0,0,50}
\definecolor{greenSimon}{RGB}{20,150,40}

\newcommand{\karman}{von~K\'arm\'an}
		
\usepackage{ulem}

\begin{document}

\title{{Suspension of large inertial particles in a turbulent swirling flow}}

\author{Benjamin Laplace}
\affiliation{Univ Lyon, Ens de Lyon, CNRS, Laboratoire de Physique, Lyon, France}    

\author{J\'er\'emy Vessaire}
\affiliation{Univ Lyon, Ens de Lyon, CNRS, Laboratoire de Physique, Lyon, France}
\affiliation{Univ. Grenoble Alpes, CNRS, Grenoble INP, Institut N\'eel, 38000 Grenoble, France}               

\author{David Oks}
\affiliation{Univ Lyon, Ens de Lyon, CNRS, Laboratoire de Physique, Lyon, France} 

\author{Oliver Tolfts}
\affiliation{Univ Lyon, Ens de Lyon, CNRS, Laboratoire de Physique, Lyon, France}    
\affiliation{Univ. Grenoble Alpes, CNRS, Grenoble INP, LEGI, 38000 Grenoble, France}   

\author{Mickael Bourgoin}
\affiliation{Univ Lyon, Ens de Lyon, CNRS, Laboratoire de Physique, Lyon, France}    

\author{Romain Volk}
\email{romain.volk@ens-lyon.fr}
\affiliation{Univ Lyon, Ens de Lyon, CNRS, Laboratoire de Physique, Lyon, France}               

\date{\today}

\begin{abstract} 
We present experimental observations of the spatial distribution of large inertial particles suspended in a turbulent swirling flow at high Reynolds number. The plastic particles, which are tracked using several high speed cameras, are heavier than the working fluid so that their dynamics results from a competition between gravitational effects and turbulent agitation.
We observe two different regimes of {suspension}. At low rotation rate, particles are strongly confined close to the bottom and are not able to reach the upper region of the tank whatever their size or density. At high rotation rate, particles are loosely confined: small particles become nearly homogeneously distributed while very large objects are preferentially found near the top as if gravity was reversed. We discuss these observations  in light of a minimal model of random walk accounting for particle inertia and show that large particles have a stronger probability to remain in the upper part of the flow  because they are too large to reach descending flow regions. As a consequence particles exhibit random horizontal motions near the top until they reach the central region where the mean flow vanishes, or until a turbulent fluctuation gets them down. 
 
 \end{abstract}

\maketitle

\section{Introduction}

Predicting the distribution of heavy particles transported in a turbulent flow is a long standing problem with applications in natural situations, such as sediment transport in rivers \cite{garcia_sediment_book}, as well as in industrial processes for which it may be important to achieve minimal agitation so that all particles are suspended in a chemical reactor \cite{zwietering,handbookmix}. From a practical point of view, the state of the suspension is given by a competition between gravity which makes particles settle and turbulent agitation which tends to suspend them. Such problem is complex as it depends not only on the size and density of the particles, but also on the properties of the flow (topology, amplitude, level of turbulence...). It was early recognized that one could draw an analogy between turbulent agitation and Brownian motion \cite{bib:Kampen}. Neglecting particles inertia, Rouse developed such analogy to derive an equation for sediment concentration profile in rivers and channel flows \cite{rouse1937,vanrijn}. Such an approach is still used to predict  particle concentration field of micron-sized particles in agitated vessels \cite{barresi1987,ayazi1989,magelli1990}, and has been refined using two phase flow models which compare well with experimental data \cite{montante2005}.

However, inertia starts to play a major impact when increasing the size of the particles or their density contrast with the carrier fluid, as shown by recent Lagrangian studies of small inertial particles dynamics in homogeneous isotropic turbulence \cite{bib:toschi2009_AnnRevFluidMech}. The response of inertial particles to turbulent fluctuations is even more complex when their size lies in the inertial range of the turbulence because such material particles are less sensitive to fluctuations at scales smaller than their size \cite{bib:qureshi2008_EPJB}, a situation encountered in many applications. In the last decade efforts have been carried out to characterize the dynamics of even larger particles whose sizes are of the order of the integral length scale of the flow \cite{Zimmermann:prl2011,Klein:mst2013,Cisse:jfm2013,machicoane_volk_pof_2016}. Their dynamics was found to be strongly influenced both by their response to local turbulence and by the global topology of the flow; they were found to exhibit preferential sampling in non homogeneous turbulent flows with back-and-forth motions between attractors corresponding to low pressure regions of the flow \cite{machicoane:njp2014,machicoane2016stochastic,machico_2021}. This preferential sampling was found to be only weakly affected by density as light and heavy particles of a given size were found to explore the same regions. All these effects were studied in a regime for which buoyancy was negligible, and one may wonder how large-and-heavy particles distribute in a situation where gravitational effects are comparable with turbulent agitation.

In the present article we address the question of the {dynamics} of large particles denser than the carrier fluid {which are suspended in a flow} and study their position probability distribution function {under gravity} when varying flow parameters in the fully turbulent regime. {The present problem is then different from re-suspension which focusses on how heavy particles can detach from the bottom as studied for instance in \cite{traugott,shnapp}}. Our experiment is performed in a swirling flow, forced with one rotating disc situated on the top, in which inertial particles are tracked optically using several high speed cameras as described in section \ref{sec:setup}. In such situation, heavy particles tend to settle due to gravity and are {maintained suspended} by the turbulent flow so that they explore non-uniformly the flow volume. After describing the flow properties in section \ref{sec:flow}, we show in section \ref{sec:distribution} that two regimes of {suspension} are observed. At low rotation rate, {gravitational effects are dominant and particles are strongly constrained to remain} close to the bottom and are not able to reach the upper region of the tank whatever their size or density. At high rotation rate, particles are loosely confined: small particles become nearly homogeneously distributed while very large {and heavy} objects are  preferentially found near the top as if gravity was reversed. Such trapping is much stronger than the one observed in a similar setup with 2 discs rotating \cite{machicoane:njp2014}. We discuss these observations in section \ref{sec:model} in light {of} a minimal model of random walk accounting for particles inertia. Investigation of the particle {density current} shows that large particles sample preferentially upper regions due to geometrical constraints{:} they are too large to reach descending flow regions which are localized in the corners {of the vessel}. As a consequence {large} particles have erratic horizontal motions near the top until they {eventually} reach the central region where the mean flow vanishes, or until a turbulent fluctuation gets them down. 

\section{Experimental setup \label{sec:setup}}

\begin{figure}
	\begin{center}
	        \includegraphics[scale=0.2]{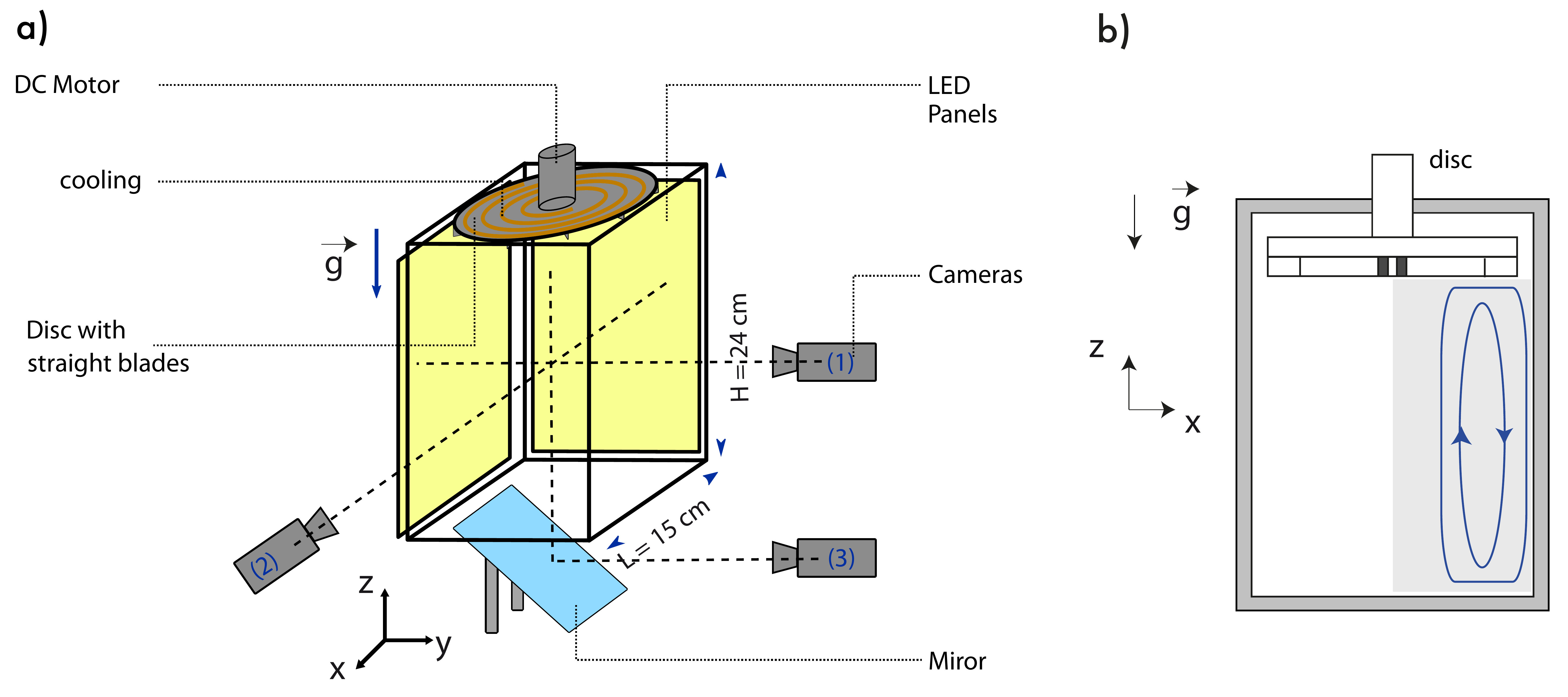}\\
		\caption{{Experimental set-up. (a) Geometry of the von K\'arm\'an flow with only one disc rotating at the top, and optical setup to perform particle tracking velocimetry.
		(b) Sketch of the mean flow topology composed of a strong azimutal motion and a poloidal recirculation.}}\label{fig:setup}
	\end{center}
\end{figure}

We study the motion of heavy particles in a \karman{} flow of water using the setup pictured in figure \ref{fig:setup}. The turbulent flow is generated in a parallelepipedic domain, with vertical length $H=24\text{ cm}$ and a square cross section with width $L = 15$ cm. The flow is produced by a disc of radius $R=7.1$ cm placed on top of the tank, fitted with 8 straight blades {with height $7$ mm}, rotating at constant frequency $\Omega$ to impose an inertial steering. The fluid is set in motion near the rotating disc so that it generates a turbulent flow with a mean flow topology resembling a vertical tornado: it is dominated by a strong mean rotation around the vertical direction $z$, which is also the direction of gravity. Because the domain has a square rather than cylindrical cross section, a strong secondary circulation takes place induced by ejection of fluid at the edge of the disc so that the mean flow is 3D with strong vertical motions. 
The working fluid is water with imposed temperature $T=20^\circ$C using a cooling plate placed at the top of the vessel so that the fluid kinematic viscosity is $\nu=10^{-6}$ m$^2$.s$^{-1}$. Operating with rotating frequencies in the range $\Omega \in [1.5,~6]$ Hz, the large scale Reynolds number of the flow, $Re = 2\pi R^2 \Omega/\nu$, is always larger than $5~10^4$. As the steering is inertial, mean flow components and velocity fluctuations are proportional to the large scale velocity $U=2 \pi R \Omega$ in this fully turbulent regime. {We estimate the {global} energy injection rate $\mathcal{I}$ {at a given rotation rate $\Omega$} using the formula {$\mathcal{I} = (P_{water}(\Omega) - P_{air}(\Omega))/M$ where $P_{water}(\Omega)$ and $P_{air}(\Omega)$ are the total electrical power consumption of the motor in water and air respectively and $M$ is the total mass of water \cite{Volk:jfm2011}}. Such global estimate of energy input contains contributions from the mean flow and from turbulent fluctuations, the latter being referred to as $\epsilon$ and called dissipation. $\mathcal{I}$ and $\epsilon$ being of the same order of magnitude in von K\'arm\'an flows, we use $\mathcal{I}$ as an estimate of $\epsilon$ and make the approximation $\epsilon=\mathcal{I}$ in the sequel.} {Results are given in  Table~\ref{table1} together with measurements of the mean vertical {and azimuthal velocities} and the fluctuating velocity, $u'$, obtained by laser doppler velocimetry at the location $(x,y,z)=(4,0,12)$ cm, where the mean flow is strong {and varies only weakly with the altitude}. Combining these measurements, it is possible to obtain the main flow parameters such as the Kolmogorov scale {$\eta=(\nu^3/\epsilon)^{1/4}$, which decreases by a factor two when increasing the Reynolds number, and the flow integral scale $L_\mathrm{int}=(u')^3/\epsilon$ which varies only weakly in the range of Reynolds number considered here.}


\begin{table}[h!]
\begin{center}
\begin{tabular}{|c |c |c |c |c |c |c |c|c|c|}
\hline	
$\Omega$ [Hz] & $u'$ [cm/s]  & $\overline{u}_z$ [cm/s] & {$\overline{u}_y$ [cm/s]} & $u_\mathrm{trms}$ [cm/s]  & $\epsilon$ [m$^2$/s$^3$] & $\eta$ [$\mu$m] & {$L_\mathrm{int}$ [cm]} & $Re$ [$1.10^5$]  \\ \hline
	$2$   &  8.6   & 10 & {28} & 13 & $0.1$ &  $56$  & {6.4} & $0.7$       \\                            
	$4$   &  14.4 & 21 & {57} & 25 & $0.6$ &  $36$  & {5.0} & $1.3$       \\    
	$6$   &  21.9 & 32 & {86} & 39 & $2.0$ &  $27$  & {5.3} & $2.0$       \\    
	\hline
\end{tabular}
\end{center}
\caption{Experimental parameters. $\Omega$: disc rotation frequency; {$u'=\sqrt{({u'_z}^2+2 {u'_y}^2)/3}$}: velocity fluctuation averaged over components as measured by Laser Doppler Velocimerty at mid height $(x,y,z)=(4,0,12)$ cm; {$(\overline{u}_z,\overline{u}_y)$: mean vertical and mean azimutal velocities measured by LDV at the same location}. $u_\mathrm{trms} = \sqrt{\overline{u}_z^2+(u')^2}$ : true root mean square velocity; $\epsilon$: energy dissipation estimated from the electrical power consumption of the motors;  $Re = 2\pi R^2 \Omega/\nu$: Reynolds number computed using the disc tip velocity.}
\label{table1}
\end{table}


In the following, we study the Lagrangian dynamics of PolyaMid (PM) and PolyaCetal (PC) spheres with respective density $\rho_\text{PM} = 1.14$~g.cm$^{-3}$ and  $\rho_\text{PC} = 1.4$~g.cm$^{-3}$, and diameters $d_p \in [2,~18]$~mm (accuracy $0.01$~mm, Marteau \& Lemari\'e, France). The particles in this study are therefore {\it inertial} as they are heavier than the fluid, and {\it large} because their diameter is much larger than the Kolmogorov scale of the flow, the largest sizes being {larger} than the flow integral length-scale. As the particles are heavier than the fluid, they settle in still water. We give in Table \ref{table2} their settling velocity, $v_s$, {measured {by releasing particles one at a time} in a $30\times 30 \times 80$ cm$^3$ column containing still water,} together with their corresponding Reynolds number $Re_s = v_s d_p / \nu$ and their Galileo number $Ga=v_g d_p /\nu$, where $v_g=\left( (\rho_p/\rho_f-1) g d_p \right)^{1/2}$ is the gravitational velocity obtained by assuming inertial forces are balanced with the apparent weight of the particle in the stationary regime \cite{uhlmann_doychev_2014}. It shall be noted that the Galileo number of the particles is always larger than $180$ so that they are not expected to fall steadily in still fluid, but exhibit a non trivial dynamics \cite{uhlmann_doychev_2014,huisman_colonne}. The settling velocity $v_s$ was therefore averaged over $30$ trajectories.

\begin{table}[h!]
\begin{center}
\begin{tabular}{|c|c|c|c|c|c|c|c|c|} 
	\hline 
	Type & $d_p$ [mm] & $d_p / \eta$ &  $v_s$ [cm/s] & $Re_s$ & $v_g$ [cm/s] & Ga \\
	\hline
	PM &  $3$  &  $50 - 110$  &   $9$ & 270 & $6.5$ & $200$\\ 
	PM &  $6$  &  $100 - 230$ &   $14$ & 840 & $9.2$ & $550$\\ 
	PM &  $10$ &  $180 - 380$ &  $16$ & 1600 & $11.8$ & $1180$\\ 
	PM &  $15$ &  $270 - 560$ &   $20$ & 3000 & $14.5$ & $2170$\\ 
	PM &  $18$ &  $320 - 670$ &   $23$ & 4140 & $15.9$ & $2860$\\ 
	PC &  $2$  &  $40 - 80$   &   $12$ & 240 & $8.9$ & $180$\\ 
	PC &  $6$  &  $100 - 230$ &   $24$ & 1440 & $15.5$ & $930$\\ 
	PC &  $10$ &  $180 - 380$ &  $30$ & 3000 & $20$ & $2000$\\
	PC &  $18$ &  $320 - 670$  & $36$ & 6480 & $26.8$ & $4830$\\  	
	\hline
	\end{tabular}
\end{center}
\caption{{Particles characteristics: PM PolyaMid particles; PC PolyaCetal particles, $d_p$ particle diameter; {$\eta \in [27,56]$ $\mu$m} Kolmogorov length-scale reported in Table \ref{table1}. $v_s$  settling velocity measured in $30\times 30 \times 80$ cm$^3$ column containing still water; Reynolds number $Re_s = v_s d_p / \nu$ based on the settling velocity; $Ga=v_g d_p /\nu$ Galileo number based on the gravitational velocity $v_g=\left( (\rho_p/\rho_f-1) g d_p \right)^{1/2}$.}}
\label{table2}
\end{table}

The study is based on the use of 3 datasets obtained with two particle densities. For the 2 first sets of experiments, the motion of PolyaMid (PM) or PolyaCetal (PC) particles is tracked in the whole flow volume using 3 high-speed video cameras (Phantom V.12, Vision Research, 1Mpix@6kHz) recording synchronously 3 views at approximately 90 degrees (figure \ref{fig:setup}). The flow is illuminated by 2 LEDs panels in a front face configuration with the horizontal cameras so that particles appear as black discs on bright images for cameras (1) and (2), and appear as bright objects on camera (3) which records images with a bottom view using an inclined mirror.  {We chose to run the experiment with an ensemble of particles with only one density but different diameters and use the apparent size of the objects to remove ambiguous position matching in order to improve particle tracking.} For each experiment with an imposed rotation rate $\Omega=[1.5, ~6]$ Hz, an ensemble of independent movies containing 8300 images is recorded with different sampling frequencies $f_{sampling}=[1,500,3500]$ Hz and a spatial resolution $1200\times 800$~pixels$^2$ covering all the measurement volume. Using pixel coordinates of all detected objects in each view, stereomatching in 3D is achieved using the algorithms described in \cite{machicoane_calib,bourgoin_rsi_calib}. We use different sampling frequencies in order to generate datasets with different temporal resolutions and improve statistical convergence. The 2 largest frame rates allow for reconstruction of particle trajectories using the algorithm described in \cite{Machicoane_tracking} for which particle velocity and acceleration can be computed, while the smallest frame rate $f_{sampling}=1$ Hz, provides nearly uncorrelated realizations where only position is resolved which improves position statistics convergence.\\
The dataset was then complemented by an ensemble of runs for which one PolyaMid particle of each size is tracked in the whole flow volume using only 2 Phantom V.10 cameras (4Mpix@400Hz) recording synchronously 5600 pairs of images (with camera 1 and 2 in figure \ref{fig:setup}). For this third dataset, $20$ movies with $\Omega=[3.3, ~3.7, ~4.3, ~4.7, ~6.0]$ Hz are recorded with a sampling frequency set to $50$ Hz in order to obtain very long trajectories where only position is resolved. {Such additional runs, with slightly different parameters, were motivated by a rapid transition observed in the distribution of PM particles position when increasing the rotation rate of the disc.}

{\section{A proxy for  the mean flow topology at very high Reynolds number}}
\label{sec:flow}

\begin{figure}[h!] 
	\begin{center}
        {\includegraphics[height=4.2cm]{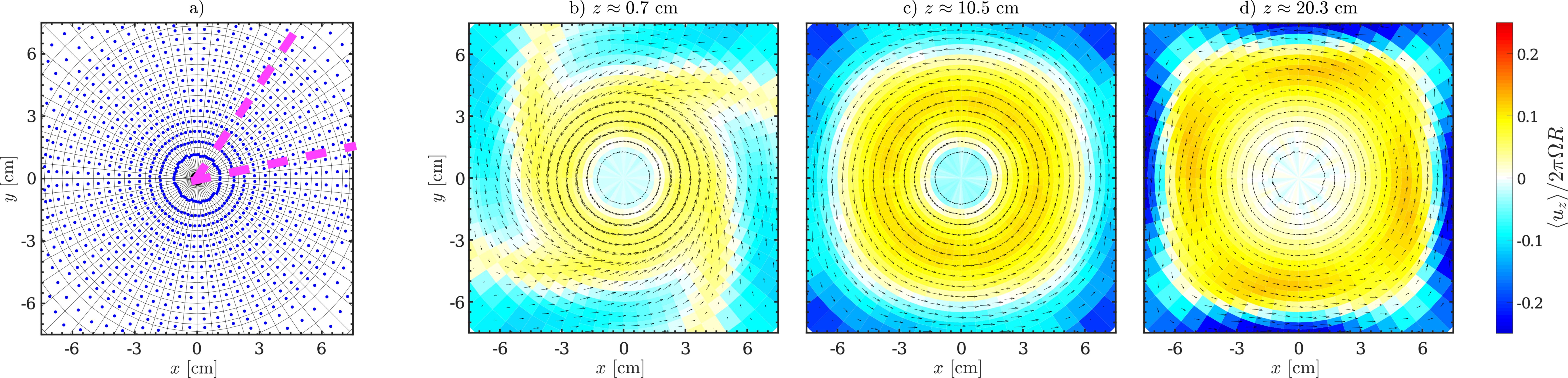}}
	\vspace{0.5cm}
	{\includegraphics[height=6.0cm]{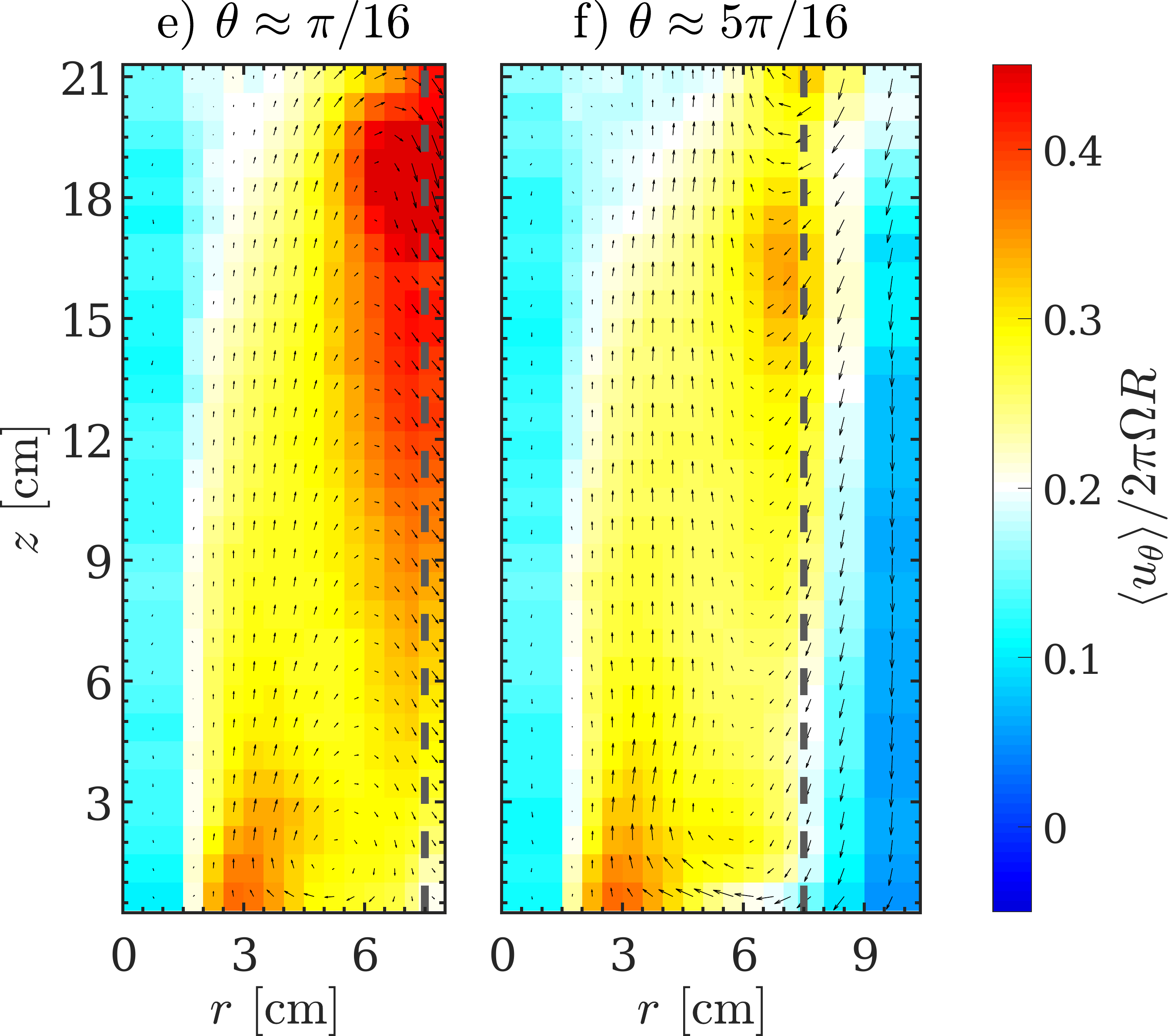}}
	\hspace{1cm}
	{\includegraphics[height=6.0cm]{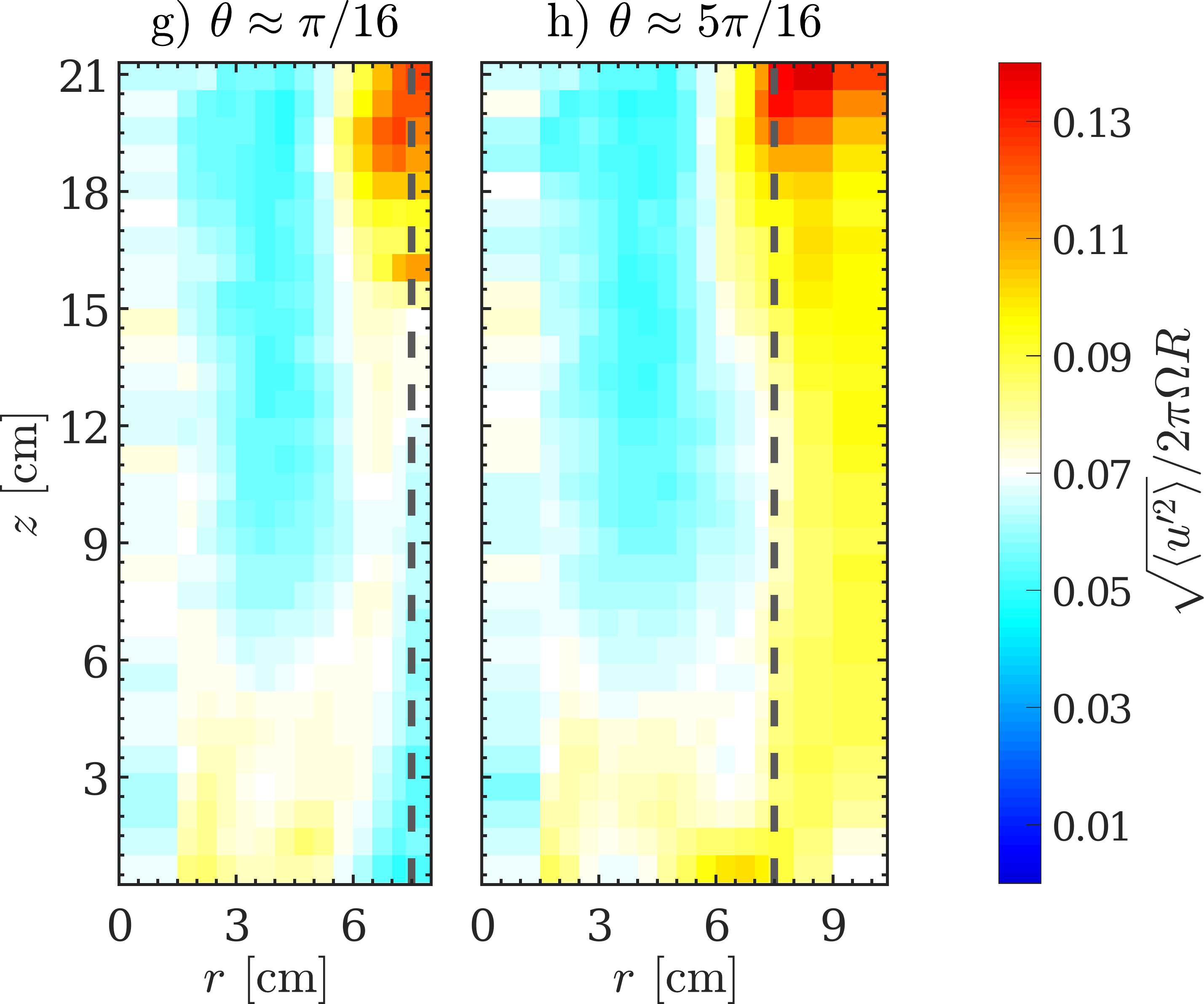}}	
	\end{center}
	\caption{a) cylindrical polar grid used to compute mean flow properties for PM3 particles. {The two dashed lines mark the two vertical planes at $\theta \approx \pi/16$ and $\theta \approx 5 \pi/16$ corresponding to figures e) and f) respectively.} Mean flow components normalized by the disc velocity $2 \pi \Omega R$ for $3$ different heights: b) $z \approx 0.7$ cm, c) $z \approx 10.5$ cm and d) $z \approx 20.3$ cm. Arrows represent a quiver plot of horizontal components ($\langle u_x \rangle, \langle u_y \rangle$), colors: vertical component $\langle u_z \rangle$. e) and f) Mean flow components in two vertical planes at $\theta \approx \pi/16$ and $\theta \approx 5 \pi/16$ respectively. Dashed lines indicate the half width of the tank, $L/2$. g) and h) Fluctuation magnitude $\sqrt{\langle u'^2 \rangle}/2 \pi R \Omega$ in the same vertical planes, with $u'^2=(u'_r)^2+(u'_\theta)^2+(u'_z)^2$ and $u'_k = u_k - \langle u_k \rangle$ ($k=r,\theta,z$).
} 
	\label{fig2:coupeuz}
\end{figure} 

When performed with sufficient time resolution, 3D particle tracking yields a set of particle trajectories each containing the temporal evolution of Lagrangian velocity $\vec V_L(t)$ at the particle position $\vec X_p(t)$. From this ensemble of {particle} trajectories, one may define a mean {particle} flow in the Eulerian framework by binning the measurement volume in small sub-volumes where the Lagrangian data can be conditioned \cite{machicoane:njp2014,huck2017production}. As the flow is expected to exhibit nearly axisymmetrical properties close to the axis of rotation ($z$), binning is performed in a cylindrical-polar grid $(r,\theta,z)$ where the size of the bins ($\Delta r$, $\Delta \theta$, $\Delta z$) is adjusted depending on the size of the particle. 

In the case of the less inertial particles (PM3: PolyaMid with $d_p=3$ mm), we used the grid depicted in figure \ref{fig2:coupeuz}a) with $N_r=17$ bins in the radial direction, $N_\theta=64$ bins in the azimutal direction, and $N_z=30$ bins in the vertical direction which corresponds to ($\Delta r \simeq 5$ mm, $\Delta \theta=0.049$ rad, $\Delta z=7$ mm). Such bin size was small enough in order to properly resolve the mean flow while ensuring good statistical convergence.

\begin{figure}[h] 
	\centering
	{\includegraphics[height=10.0cm]{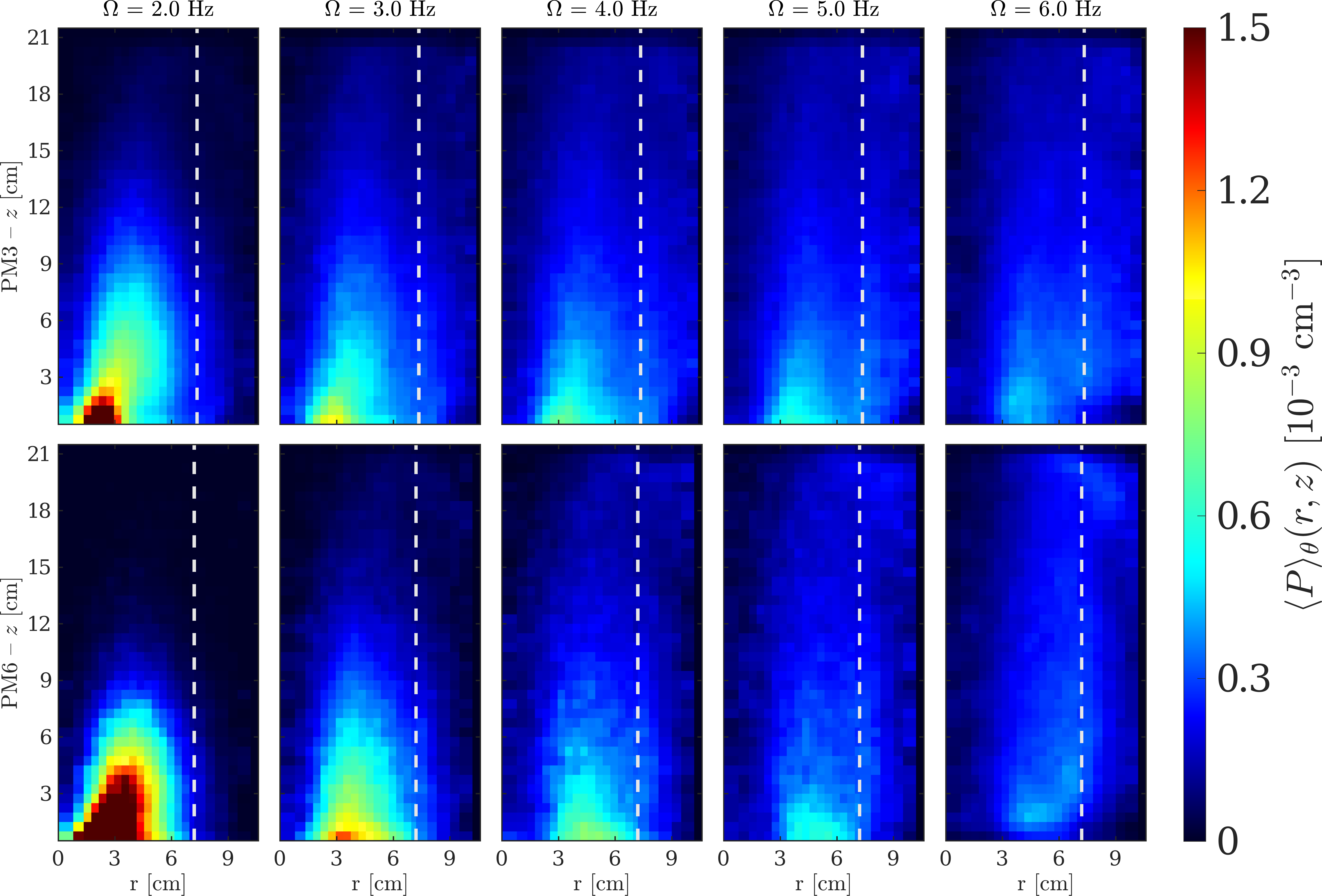}}
\caption{{Axisymetric PDFs of position for PM3 (first line) and PM6 (second line) particles plotted with the same colorbar. From left to right, rotation frequencies are $\Omega = 2.0$, $3.0$, $4.0$, $5.0$ et $6.0$ Hz. Dashed lines correspond to the bulk limit accessible to the particles: $r = L/2-d_p/2$.}} \label{fig3_cartes}
\end{figure}

As the particles are heavier than the fluid, the velocity field reconstructed from the particle trajectories is expected to be heavily biased by gravitational effects if the typical flow velocity is of the order of the particle settling velocity. In order to get insight into the topology on the flow field, we shall only focus on the mean flow reconstructed from the dynamics of PM3 particles at the highest rotation frequency $\Omega=5$ Hz, which correspond to a true root mean square velocity of the flow $u_{trms} = \sqrt{u'^2 + \overline{u}_z^2} \simeq 3.7 v_s$. In this regime, we observed that both the mean flow and fluctuations fields are found proportional to the disc velocity $2 \pi R \Omega$ as observed in \cite{machicoane:njp2014}, confirming that {particle settling has only a weak impact on the mean flow reconstruction.} Figure \ref{fig2:coupeuz}b-f displays the reconstructed {particle} mean flow in three horizontal planes (bottom, mid height, top) and two vertical planes at $\theta=\pi/16$ (close to the plane $y=0$) and $\theta=5 \pi/16$ (main diagonal direction). As can be observed, the mean flow is composed of a strong mean rotation imposed by the disc around the vertical direction, and a secondary circulation with an intense vertical component directed toward the disc for $r \leq 6$ cm, and negative in the corners where the downward motions concentrate. However, as opposed to classical von K\'arm\'an flows produced with only one disc rotating \cite{ravelet2004}, we find here that the mean flow vertical component is vanishingly weak in a nearly conical region with radius $r \simeq 1.5$ cm close to $z=0$, and $r \simeq 2.5$ cm near the top. The presence of this core is not an artefact due to the particles inertia, and was confirmed by complementary measurements performed in a sub volume using polystyrene particles with diameter $800$ $\mu$m and density {$1.06$ g.cm$^{-3}$}. This dataset showed a good agreement with the reconstructed mean flow from the PM3 trajectories {despite the fact that PM3 particles are strongly inertial as they are heavier than the fluid and much larger than the Kolmogorov length-scale of the flow. We shall use this mean flow in the sequel as a proxy for the mean flow to interpret the particle distribution.}

\section{Spatial distribution of particles}
\label{sec:distribution}

As the particles are heavier than the fluid, a minimum speed {(dependent on particle mass, size and density)} is expected to maintain them suspended in the flow \cite{zwietering}. {Intuitively one expects that particles get more suspended (i.e. their distribution become more uniform) as the rotation rate is increased, and that this tendency requires larger rotation rates for heavier particles (i.e. for larger particles at fixed density). {This is because weight and Archimede's force grow as $d_p^3$ while drag increases at most as $d_p^2$ in the non linear regime.} In this section, we investigate this scenario in the case of PolyaMid (PM) particles by exploring their spatial distribution as a function of the rotation rate for various particle diameters. We therefore compute the stationary PDF of the particle positions $P(r,\theta,z)$ in cylindrical polar coordinates, and show results for the axisymmetric part of the PDFs, denoted $\langle P \rangle_\theta=\sum_i V_i P(r,\theta_i,z)/\sum_i V_i$, where $V_i$ is the volume of the bin  located at $(r,\theta_i,z)$.} 

Results are displayed in a half cross section of the vessel in figure \ref{fig3_cartes} and figure \ref{fig4_cartes}, with the disc on the upper part (at $z = 22$ cm) and the rotation axis $(r=0)$ placed on the left of each map. We also represent the limit $r_\mathrm{lim}=L/2-d_p/2$ as a white dashed line so that the portion of the PDF with $r \leq r_\mathrm{lim}$ corresponds to the bulk part of the flow while $r \geq r_\mathrm{lim}$ corresponds approximately to the corners of the vessel.

Let us first start by commenting the evolution of $\langle P \rangle_\theta(r,z)$ in the case of the smallest PolyaMid (PM3 and PM6) particles which are displayed in figure \ref{fig3_cartes} for $\Omega=[2,3,4,5,6]$ Hz. In the case of the smallest rotation rate ($\Omega=2$ Hz), the particles explore the flow in a very heterogenous manner and their position PDFs are decreasing functions of $z$ {with} a maximum in the vicinity of $r=2-3$ cm depending on their diameter. In this  {strongly constrained regime, for which the true root mean square ($\mathrm{trms}$)} value of the fluid velocity is at most of the order of the settling velocity of the PM3 particles, the shape of the PDFs is very strongly linked to the flow topology. This is because when the particles are located near the bottom, they are advected toward the center by the radial recirculation (figure \ref{fig2:coupeuz}b)) until they reach the region where the vertical flow velocity is maximum, which happens at $r=2-3$ cm depending on the altitude as shown in figure \ref{fig2:coupeuz}e,f). As a consequence, the PDFs appear distorted following the streamlines of the velocity field, and they present a nearly conical region close to $r=0$ where the probability of finding particles is almost zero because the mean vertical velocity of the fluid is vanishingly weak in that region of the vessel. 

\begin{figure}[h] 
	\hspace{-0.39cm}
	{\includegraphics[height=10.0cm]{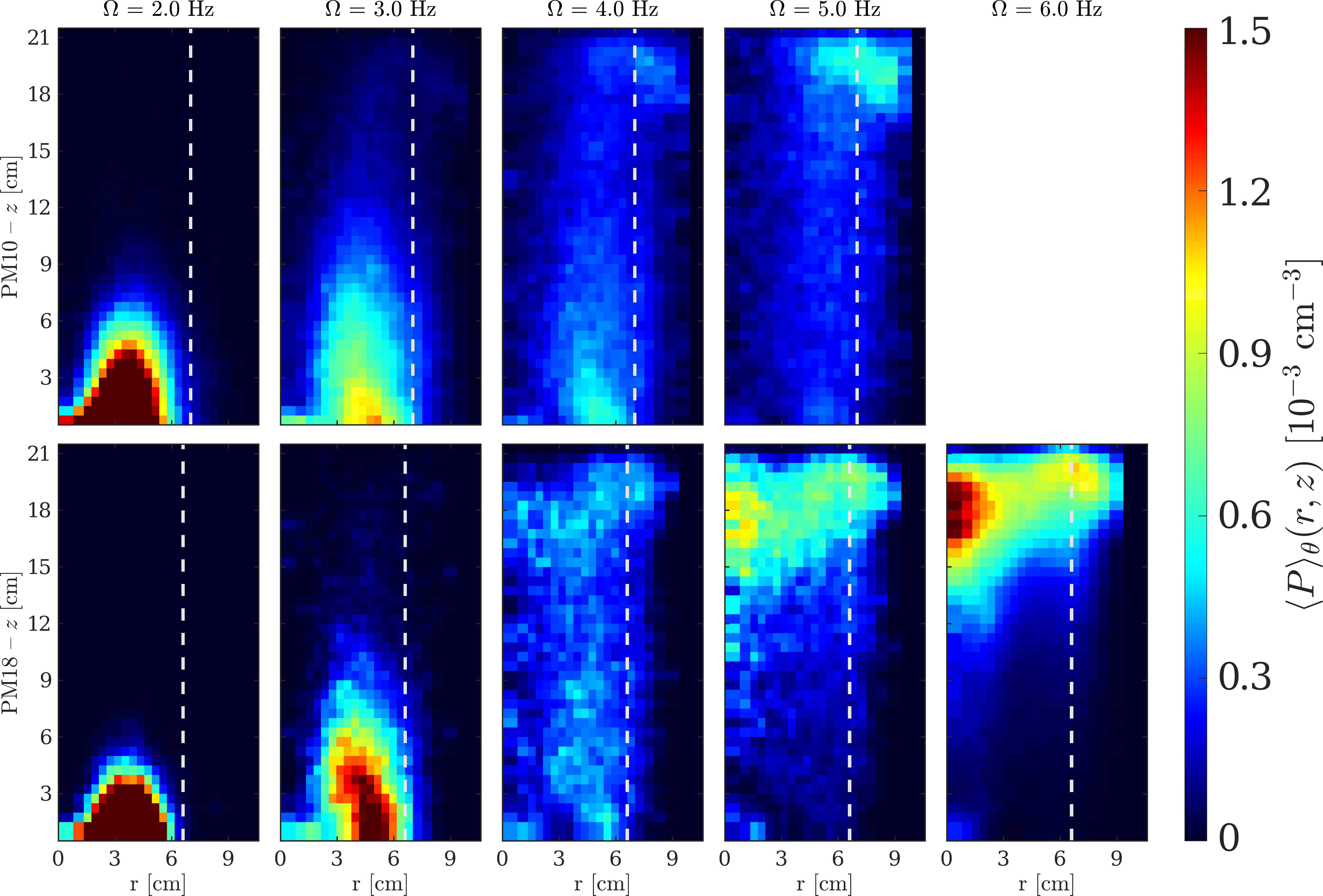}}
	\caption{{Axisymetric PDFs of position for PM10 (first line) and PM18 (second line) particles plotted with the same colorbar. From left to right, rotation frequencies are $\Omega = 2.0$, $3.0$, $4.0$, $5.0$ et $6.0$ Hz. Dashed lines correspond to the bulk limit accessible to the particles: $r = L/2-d_p/2$. No data are available in the case $\Omega=6$ Hz and $d_p=10$ mm as the particle was trapped in the spacing between the disc and one corner of the vessel, whose width almost exactly fits the particle size.}} \label{fig4_cartes}
\end{figure}

When increasing the rotation rate to $\Omega=3$ Hz, which corresponds to a fluid velocity of the order the settling velocity for PM6 particles, particles are more easily suspended so that PM3 and PM6 particles can reach the top of the vessel with a {higher} probability. The position PDFs of the particles are still tilted following the flow field, with a maximum in the radial direction located at slightly larger radii as compared to the case $\Omega=2$ Hz. This effect, which seems to be more important when increasing the particle size traces back to a possible centrifugal mechanism which will be investigated in more details further in the article. In the case of these small particles, increasing the rotation rate of the disc beyond $\Omega=3$ Hz does not affect much the position PDFs which still look similar to the one obtained at lower rotation rates although their decrease with the altitude is less and less visible when increasing $\Omega$ so that their distribution is nearly uniform at the highest rotation rates. Finding nearly homogeneous distributions in this loose confinement regime, for which gravity is less and less important, explains why the mean and fluctuating flow reconstructed from PM3 trajectories are found proportional to the disc velocity $2 \pi R \Omega$ as pointed in the previous section.

We now comment on the evolution of $\langle P \rangle_\theta(r, z)$ in the case of the largest PolyaMid (PM10 and PM18) particles which are displayed in figure \ref{fig4_cartes} for $\Omega=[2,3,4,5,6]$ Hz {and which are strikingly qualitatively different from those of smaller particles previously described}. As opposed to the previous case (PM3 and PM6) we observe now two different regimes of {suspension} depending of the rotation rate. In the case of moderate rotation rates ($\Omega \leq 3$ Hz), for which the true root mean square velocity of the fluid is smaller than the PM10 settling velocity, particles are again found preferentially close to the bottom. In this strong confinement regime, for which particles can not be transported close to the top of the vessel, their position PDFs are very similar to the one described in the case of the small particles at low rotation rates, the major difference being that the radial position at which the probability is maximum is an increasing function of the particle size. The second regime is observed for rotation rates larger than $\Omega = 4$ Hz, for which all PM particles are well suspended in the flow and have a reasonably high probability of reaching the top of the vessel. In this {loosely constrained} regime, we find that the position PDFs of particles larger than $10$ mm are almost uniform around $\Omega=4$ Hz, {while we observe for higher rotation rates} a higher probability of finding the particles close to the top. 

\begin{figure}[h!] 
	\centering
	{\includegraphics[height=7cm]{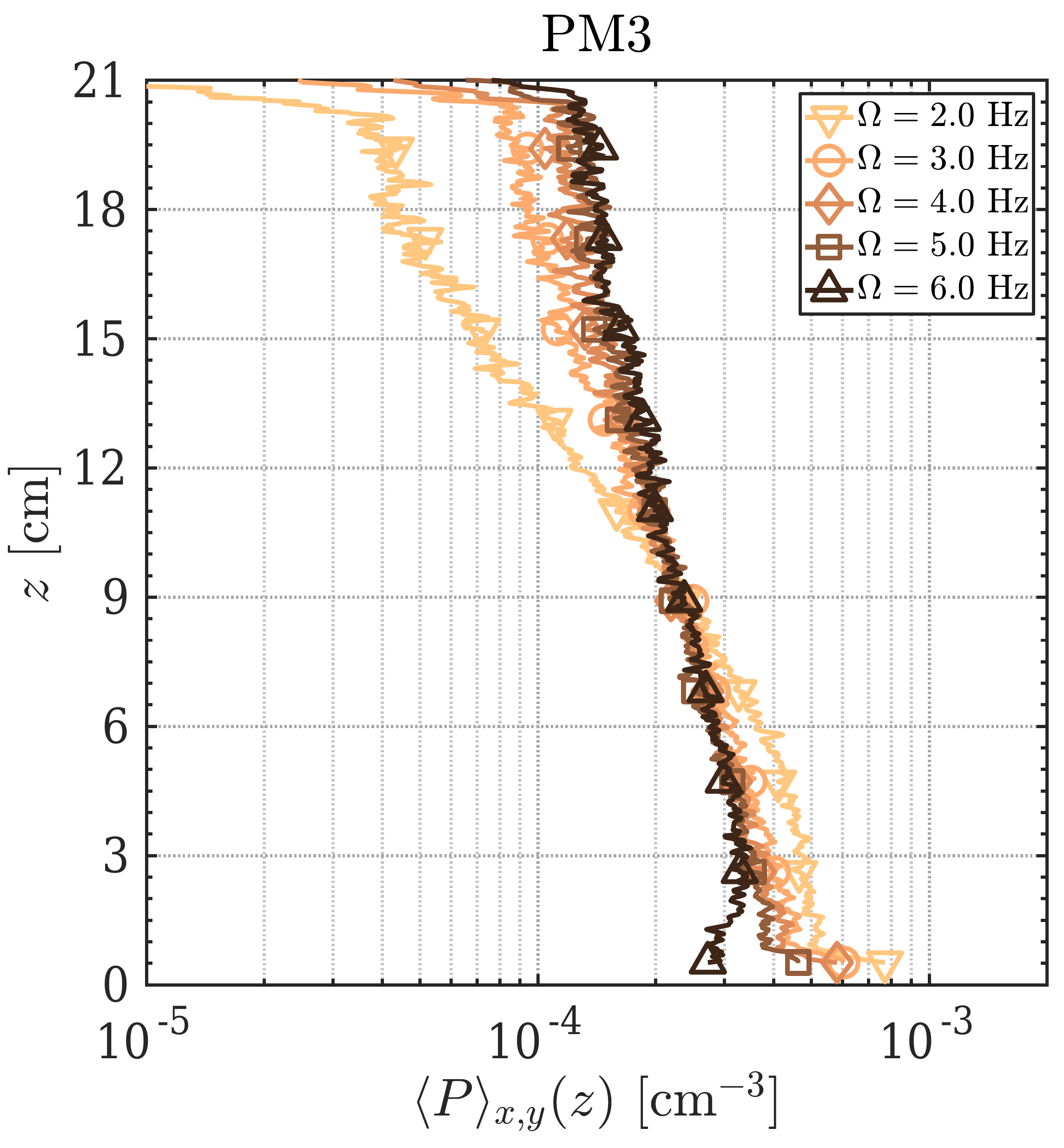}}
	\hspace{0.5cm}
	{\includegraphics[height=7cm]{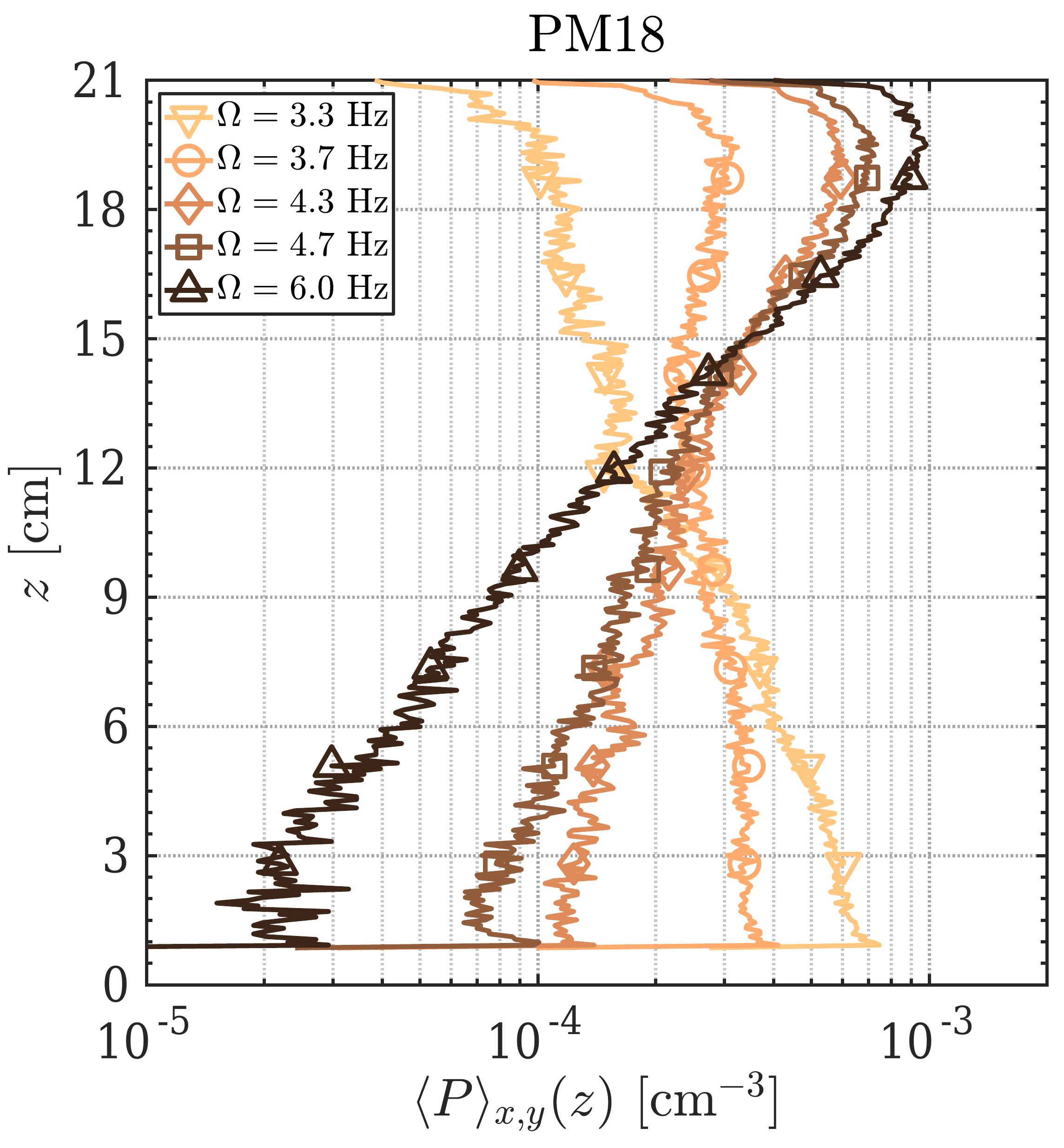}}
\caption{{1D PDFs of position, $\langle P \rangle_{x,y}(z)$, for PM3 (left) and PM18 (right) particles and different rotation rates.}} \label{fig5_pdfz}
\end{figure}

Such different behaviors between small and large particles is easily seen when looking at 1D position PDFs, $\langle P \rangle_{x,y}(z)$, obtained by averaging 3D PDFs over $x$ and $y$. Figure \ref{fig5_pdfz}a,b) display these functions {respectively} for the smallest (PM3) and largest (PM18) particles, at increasing rotation rates. One can observe that the 1D position PDF of small particles tends to reach an exponential shape of the form $\langle P \rangle_{x,y}(z) \propto \exp(-z/z^*)$ at high rotation rate with a positive characteristic length $z^*$ which increases with $\Omega$. As opposed to this behavior, the PDF of the largest PolyaMid particles has a maximum close to the bottom below $\Omega=4$ Hz and close to $z=20$ cm when $\Omega \geq 4$ Hz. Assuming $\langle P \rangle_{x,y}(z)\propto \exp(-z/z^*)$ is nearly exponential in the range $z \in [3,18]$ cm, one would find that $z^*$ is positive at low rotation rate, reaches infinity around $\Omega \simeq 4$ Hz before becoming again finite but negative at higher rotation rates as if the gravity was reversed. {Such counterintuitive observation that heavier objects are more easily suspended at large rotation rate may be thought to be linked to the trapping of large particles observed in the case of other von K\'arm\'an flows, for which particles were preferentially found in the low pressure low fluctuations regions of the flow \cite{machicoane2016stochastic,machico_2021}. However the present trapping is much more intense as it is nearly $60$ times more probable to observe a PM18 particle near the top than near the bottom at the highest rotation rate, while the probability was only increased by a factor 2-3 in \cite{machicoane2016stochastic,machico_2021}. The mechanisms at work for the dynamics of small and large objects will be discussed in the next section.}

{\section{Evolution of mean height and mean radial position}}
\label{sec:model}

The 2D position PDFs being functions of $(r, z)$, {we sum up all the observations for the sake of comparing trends with the rotation rate and the different configurations and plot} in figure \ref{fig5:hmean} the mean altitude $\overline{z}/H$ and mean {radial position} $\overline{r}/(L/2)$ as a function of $\Omega$ for each particle type (PolyaMid and PolyaCetal). As it could have been anticipated, $\overline{z}$  is an increasing function of the rotation rate because the forcing from the flow increases with $\Omega$, however PolyaCetal particles {are found to be more difficult to suspend,} even at $\Omega=6$ Hz due to their higher density. Indeed, at the maximal velocity, the {true root mean square} value of the {flow} velocity is only of the order of the settling velocity of large PC particles so that they always remain in the {strongly constrained} regime and never reach the top of the vessel. Such result was confirmed by looking at their 2D position PDFs (not shown here). Suspending these particles would have required to increase the rotation rate beyond $6$ Hz, which may have damaged the setup due to particle collisions against the lateral walls. Indeed, the average distance to the axis of rotation, $\overline{r}$, is also an increasing function of the rotation rate (figure \ref{fig5:hmean} b)), at least when particles are in the {strongly constrained} regime. {Indeed, the largest PM particles exhibit a mild decrease of the mean radial position beyond, in the loosely constrained regime, when their 2D position PDF present a maximum in the upper part of the vessel.}

\begin{figure}[h] 
	\centering
	{\includegraphics[height=7.0cm]{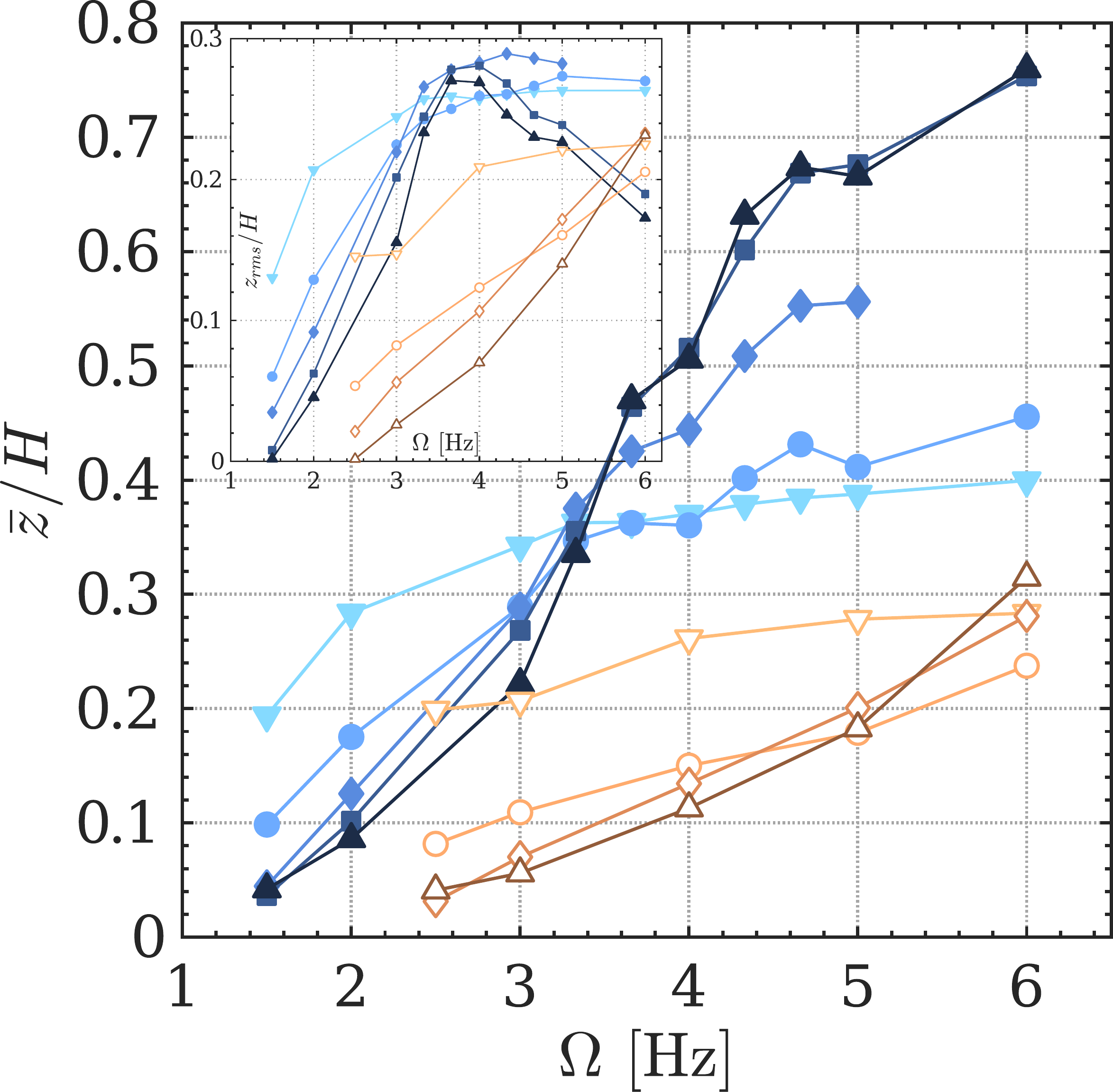}}
	\hspace{0.5cm}
	{\includegraphics[height=7.0cm]{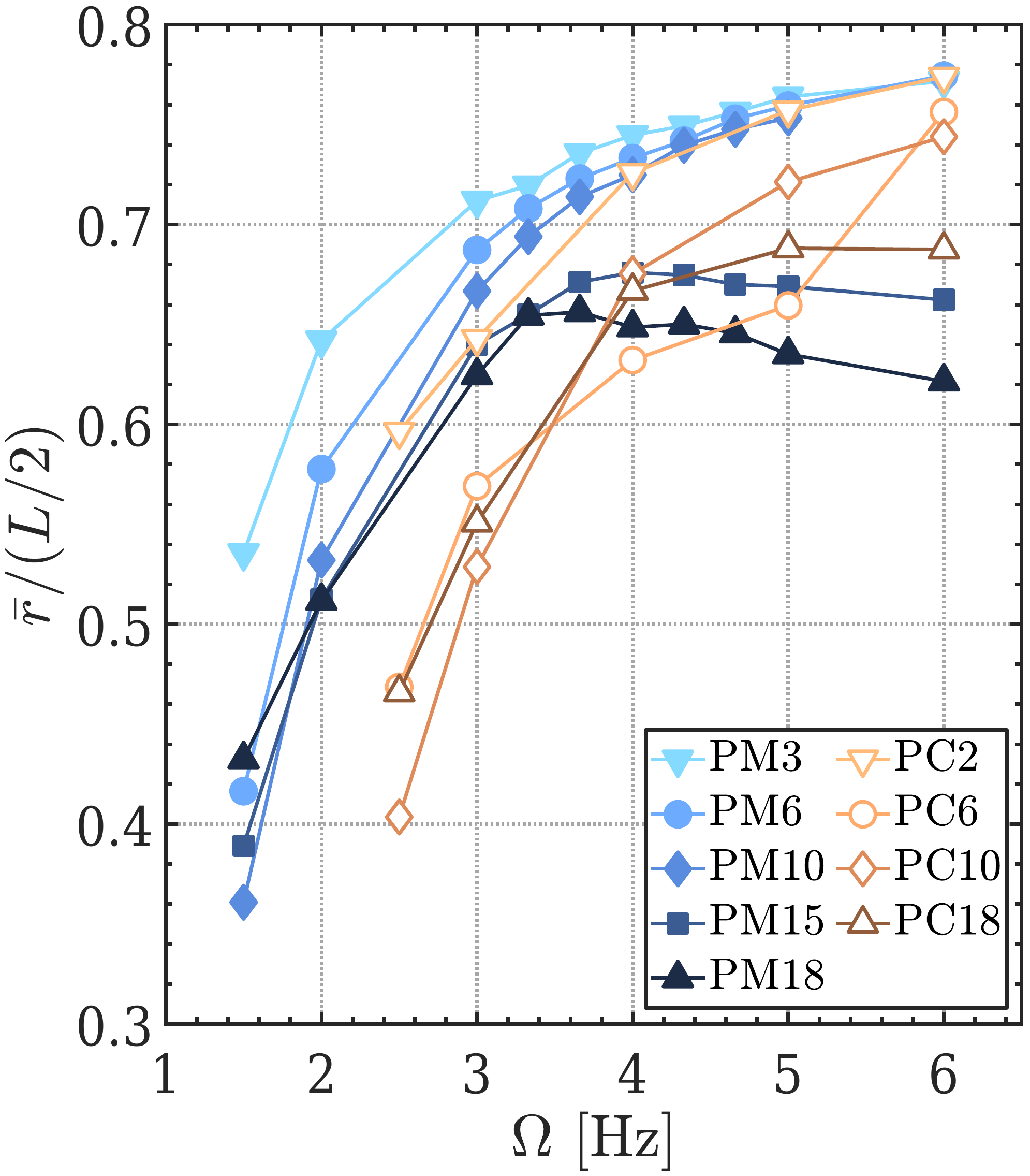}}
	\caption{ Mean normalized altitude $\bar{z}/H$ (left) and mean normalized {radial position} $\bar{r}/(L/2)$ (right) of PM and PC particles plotted against the rotation rate $\Omega$. PM3: ($\blacktriangledown$), PM6: ($\bullet$), PM10: ($\blacklozenge$), PM15: ($\blacksquare$) and PM18: ($\blacktriangle$). PC2: ($\triangledown$), PC6: ($\circ$), PC10: ($\lozenge$) and PC18: ($\triangle$). {Inset of left figure : normalized rms altitude $z_\mathrm{rms}/H$ against the rotation rate $\Omega$ for the same cases. In all cases, the error on the estimation of the mean altitude is smaller than $E=2$ mm ($E/H < 0.01$).}} \label{fig5:hmean}
\end{figure}

{Most} of the results shown here can be understood in the framework of stochastic modeling of turbulent diffusion for which one usually describes the turbulent transport of particles as a Brownian {process} . Such approach, inspired from molecular diffusion, has been widely used to model sediment transport in rivers \cite{rouse1937}. Neglecting the correlation of the turbulent flow, one obtains a zeroth order model \cite{bib:sawford1991_PoFA} for which the increment of position of an inertialess particle is:
 
\begin{equation}
d \vec X = \vec v_s dt + \langle \vec u \rangle dt + \sqrt{2 D_\mathrm{turb}} d \vec W, 
\end{equation}

where $\vec v_s$ is the settling velocity of the particle, $\langle \vec u \rangle$ the mean flow field, $\vec W$ a Wiener process \cite{bib:Kampen}, and $D_\mathrm{turb}$ a diffusion constant accounting for the contribution of turbulence. In homogeneous isotropic turbulence its expression is related to the amplitude of velocity fluctuations and the Lagrangian correlation time $T_L$ by the expression $D_\mathrm{turb}=u'^2 T_L$ \cite{bib:popebook,bib:tennekes_book}. As a consequence, the stationary PDF of particle position, $P_s(\vec x)$, obeys the Fokker Planck equation \cite{bib:Kampen}:

\begin{equation}
 \vec \nabla \cdot (\vec v_s + \langle \vec u \rangle) P_s = D_\mathrm{turb} \nabla^2 P_s.
\end{equation}

This equation is an other version of the turbulent transport equation derived using a continuous media approach for the concentration of sediments in rivers and channel flows \cite{rouse1937,vanrijn,garcia_sediment_book}. It has been widely used in 1D to predict concentration profiles in mixers \cite{barresi1987,ayazi1989,magelli1990}.

In the absence of a mean velocity field, {for the case of} heavy particles ($\vec v_s = - v_s \vec e_z$), {the problem is identical to the Jean Perrin problem of colloidal suspension in a gravitational field \cite{perrin_acp,perrin_jpta}, where molecular diffusion is replaced with the turbulent diffusion constant. The} solution with zero flux at the bottom of the tank is $P(z) \propto \exp(-z/z^*)$ with $z^*=D_\mathrm{turb}/v_s$ so that $\bar{z} /H$ is a growing function of the inverse of the turbulent P\'eclet number $Pe_\mathrm{turb}=v_s H/D_\mathrm{turb}$.

Given that 1D PDFs of small particles were close to exponential functions (figure \ref{fig5_pdfz}), one may plot figure \ref{fig5:hmean} a) as a function of $1/Pe_\mathrm{turb}$. Before doing so, it shall be noted that when the mean flow is present, $P_s$ has a 3D structure which reflects the mean flow geometry so that the problem is governed by $Pe_\mathrm{turb}$ and a second non dimensional number which we define as a second P\'eclet number based on the mean flow velocity, $Pe_U=HU/D_\mathrm{turb}$. In the present case, measurements are performed in the fully turbulent regime for which $U \propto R \Omega$, $u' \propto R \Omega$, $T_L \propto 1/\Omega$ \cite{machicoane_volk_pof_2016} so that $1/Pe_\mathrm{turb} \propto R^2 \Omega/H v_s$ {while} $Pe_U \propto H/R$ remains constant when varying the rotation rate.

Figure \ref{fig6:hmean} displays $\overline{z}/H$ as a function $1/Pe_\mathrm{turb} = R^2 \Omega/H v_s$. It is visible that the various curves are now relatively well rescaled, at least in the {strongly constrained} regime, which corresponds {to} high values of $Pe_\mathrm{turb}$ {\it i.e.} when settling dominates over {agitation}. It can also be seen that the curves separate in two bundles: whatever their density, particles with diameters larger than $10$ mm do not show any sign of saturation {around $\overline{z}/H \simeq 1/2$ (homogeneous distribution) but keep increasing up to $\overline{z}/H \simeq 1$ (particles preferentially at the top)} when $1/Pe_\mathrm{turb}$ increases while the mean height of smaller particles tend to saturate to a value smaller than $H/2$ which would be obtained for a homogeneous distribution. Moreover, this saturation value depends on the particle size and density which shows that some effects, such as particle inertia or finite-size effects, need to be taken into account to explain the mean height evolution. 

\begin{figure}[h] 
	\centering
	{\includegraphics[height=7.0cm]{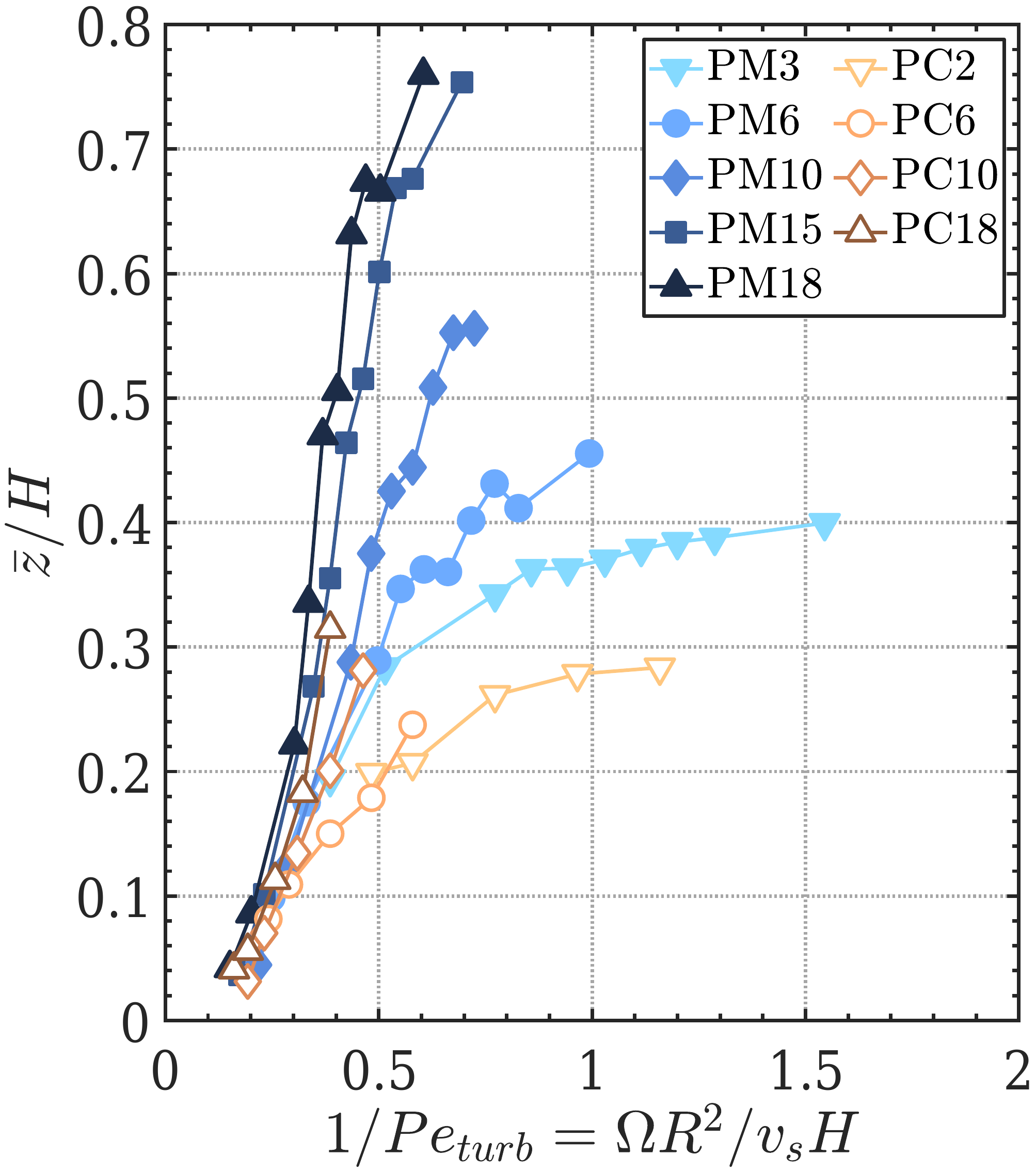}}
	\caption{{Mean normalized altitude $\bar{z}/H$ of PM and PC particles plotted against the inverse of the turbulent P\'eclet number $1/Pe_\mathrm{turb} = R^2 \Omega/H v_s$. PM3: ($\blacktriangledown$), PM6: ($\bullet$), PM10: ($\blacklozenge$), PM15: ($\blacksquare$) and PM18: ($\blacktriangle$). PC2: ($\triangledown$), PC6: ($\circ$), PC10: ($\lozenge$) and PC18: ($\triangle$).}} \label{fig6:hmean}
\end{figure}

In order to evaluate the influence of inertia, we shall {consider} the forces driving particle motions. {A reasonable start is the use of} the Maxey-Riley-Gatignol equation derived for small inertial particles transported by a flow \cite{Maxey:pof1983,Gatignol:1983}. In the case of a spherical particle of radius $a$ and density $\rho_p$, with position $\vec X(t)$, moving with a velocity $\vec V$ in a fluid of density $\rho_f$, kinematic viscosity $\nu$, and velocity field $\vec u$, the equation of motion reads:
\begin{equation}
 \frac{d \vec V}{dt} = \beta \frac{D\vec u}{Dt} + \frac{1}{\tau_p} (\vec u - \vec V) + \delta \vec g, \hspace{1cm} \frac{d \vec X}{dt} = \vec V
\end{equation}
where $D\vec u/Dt$ is the fluid acceleration, $\beta=3 \rho_f/(2 \rho_p + \rho_f )$, $\tau_p=a^2/(3 \beta \nu)$ is the response time of the particle, and $\delta  = (\rho_p-\rho_f)/ (\rho_p + \rho_f/2 )$ quantifies buoyancy. This equation, in which we have neglected the history force, is only valid in the limit of vanishingly small particles provided their local Reynolds number $Re_p = a \|\vec V-\vec u\|/\nu$ is very small. {In the present study} the particles are much larger than the smallest length-scale of the flow and the particles settle in still water at large Reynolds number so that the drag force is not expected to be linear. We shall therefore only use this toy model as an eye guide to understand the main observations. In the case of moderate inertia, for which the acceleration of the particles is close to the one of the fluid, the particle velocity {can be approximated as}:

\begin{equation}
 \vec V \simeq \vec u + \vec v_s + \tau_p (\beta-1) \frac{D\vec u}{Dt}
\end{equation}

\noindent where $\vec v_s = \tau_p \delta \vec g$ is the settling velocity, and the last term stands for the first correction due to particle inertia. In order to derive  a stochastic model valid for such case, one must split the fluid contributions into mean and fluctuating parts. Assuming turbulence is homogenous (which leads to a constant diffusion coefficient), one has $\langle {D\vec u}/{Dt} \rangle = \langle \vec u \rangle \cdot \nabla \langle \vec u \rangle$ so that the motion of the particle obeys the random walk: 

\begin{equation}
  d \vec X = \vec v_s dt+\langle \vec u \rangle dt + \tau_p (\beta-1)\langle \vec u \rangle \cdot \nabla \langle \vec u \rangle dt  + \sqrt{2 D_\mathrm{turb}} d \vec W.
\end{equation}

When taking into account inertia, particles follow an effective compressible flow which increases their probability of reaching certain regions of space. Such effect explains {the centrifugation process by which} the mean {radial position} of the particles increases with increasing the rotation rate as observed in figure \ref{fig5:hmean} b). 

Indeed, in the present case the mean flow presents a large azimutal recirculation which may be approximated as $\langle u_\theta \rangle \simeq A r \Omega$  ($A \simeq 2.5$ from figure \ref{fig2:coupeuz}) which turns into a mean radial acceleration:
\begin{equation}
\langle \vec u \rangle \cdot \nabla \langle \vec u \rangle \simeq - \frac{\langle u_\theta \rangle^2}{r} \vec e_r = - A^2 r \Omega^2 \vec e_r,
\end{equation}
such that heavy particles ($\beta-1 \leq 0$) are centrifuged out from the vortex. Because this term has a quadratic contribution to the position increment (in terms of the parameter $\Omega$), it is negligible at low rotation rate and becomes important when the Froude number $Fr=R\Omega^2/g$, which is the third number upon which the particle position PDF depends, becomes of order unity ($Fr=0.25$ for $\Omega=6$ Hz). This radial exploration has important consequences because particles of different size and density do not explore the mean flow {in the same way}. Indeed, a particle with no radial segregation will explore both ascending regions $r \in [2,~6]$ cm and descending regions located in the corners ($r \geq 7$ cm) while segregation will increase the probability of visiting ascending or descending regions depending of the particle size and density.

\begin{figure}[h!] 
	\centering
	{\includegraphics[height=6.0cm]{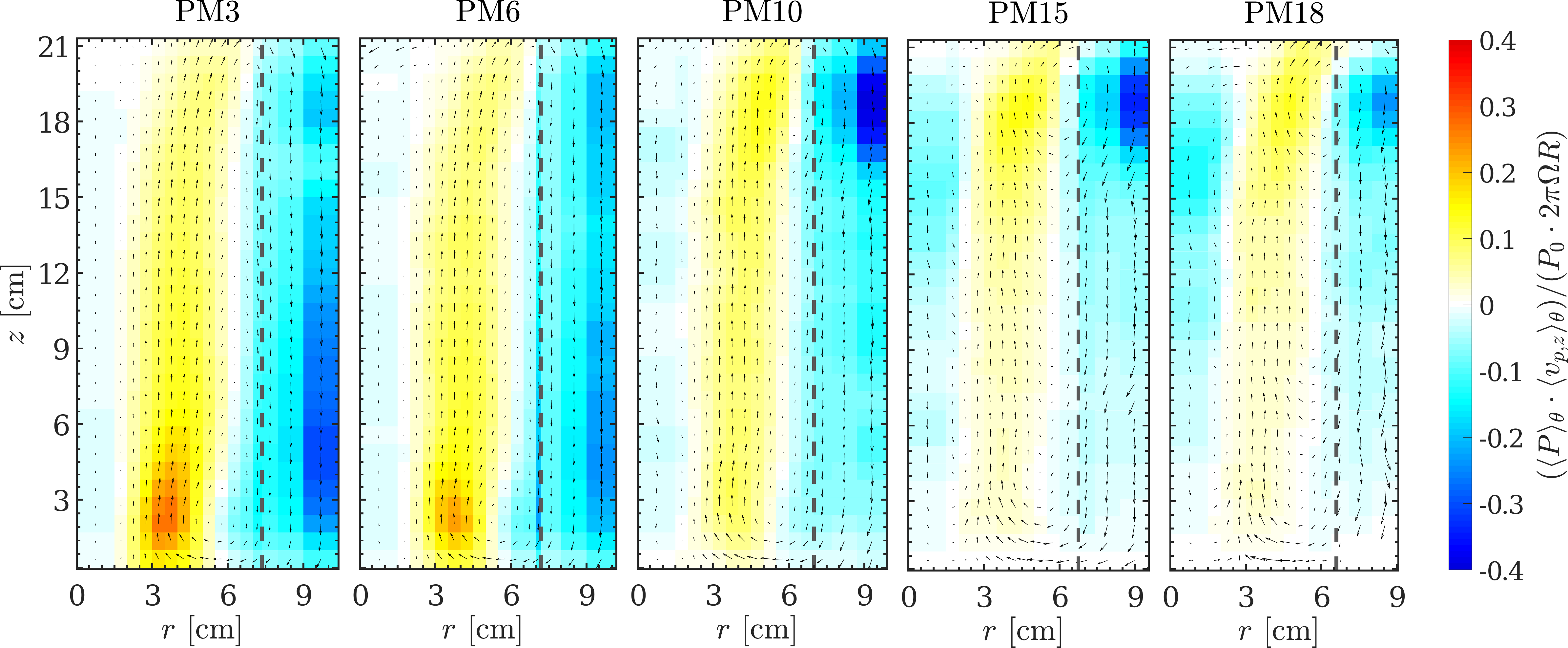}}
	\caption{{Normalized axisymetric vertical probability current $\langle P \rangle_{\theta}(r,z) \langle v_{p,z} \rangle_{\theta}(r,z)/(P_0 \cdot 2 \pi R \Omega)$, with $P_0=\langle P \rangle_{r,\theta,z}$, for PolyaMid particles and a rotation rate $\Omega=5$ Hz. From left to right PM3, PM6, PM10, PM15, PM18 particles. The domain corresponds to the volume actually accessible to each particle. Dashed lines correspond to the bulk limit accessible to the particles: $r = L/2-d_p/2$.}} \label{fig8:flux}
\end{figure}

This is visible in figure \ref{fig8:flux}, where we have plotted the vertical probability current $\langle P \rangle_\theta \langle v_{p,z} \rangle_\theta$ for PolyaMid particles and a large rotation rate $\Omega=5$ Hz, so that the mean flow {(reconstructed from the trajectories of each particle type)} is larger than the settling velocity of the particles. This figure shows that small particles are lifted up in a large portion of the tank and recirculate near the corners so that the descending current is located in the corners, where the descending velocity is the strongest. We believe this is the main reason why the mean height of small particles saturate at a smaller value than $H/2$ when the turbulent P\'eclet number goes to zero {as centrifugation tends to bias particles dynamics toward the downward regions}. 

This hypothesis was tested numerically in 2D in a finite domain $\vec X = (x,z) \in [-L/2,L/2] \times [0,H]$ where we have simulated the Brownian motion of particles using the model : 
\begin{equation}
d \vec X = (- v_s \vec e_z + \vec u + \alpha x u_0^2 \vec e_x) dt + \sqrt{L  u_0} d \vec W,
\end{equation}
where $v_s$ is the (downward) settling velocity, and $\vec u = u_0 (-\sin(2 \pi x/L) \cos(\pi z/H), 2 \cos(2 \pi x/L)\sin(\pi z/H) H/L) $ is a divergence free flow with amplitude $u_0$ with vertical upward motion near $x=0$ and downward motion near $|x| \simeq L/2$. Here $D_\mathrm{turb}=L u_0/2$ is the turbulent diffusion coefficient, $\alpha x u_0^2 \vec e_x$ accounts for the centrifugal term which pushes heavy particles on the sides when $\alpha >0$, and we use reflective boundary conditions for the position when a particle leaves the domain at some time step which is equivalent to a no flux boundary condition for the probability current \cite{bib:Kampen}. It was observed that, for $\alpha=0$ (no inertia), all the curves $\bar{z}/H$ saturate at the expected value $1/2$ at high value of $u_0$ with perfect rescaling when plotting the results as a function $Pe=u_s L/2 D_\mathrm{turb}=u_s/u_0$. In the case $\alpha >0$, the probability of reaching downward regions on the sides was increased, leading to decreased saturation values at increasing $\alpha$ so that the rescaling was only possible at small values of $u_s/u_0$.

If such a point particle approach may give an answer for small particles, this does not explain the evolution of the mean height in the case of very large particles (figure \ref{fig6:hmean}) which exhibit an increased probability of being close to top of the vessel, so that $\overline{z}/H$ reaches saturation values close to $1$ at high rotation rate. Indeed, those particles are also centrifuged out when increasing $\Omega$, although their mean radial position seems to saturate at lower values {than those of smaller particles} (figure \ref{fig5:hmean} b)). Again, one may investigate the probability current displayed in figure \ref{fig8:flux} to understand the dynamics of these large objects. {As} opposed to the small particles case, the downward flux of large particles {is found to} decrease strongly in the corners when increasing the particle diameter from $10$ to $18$ mm. {A possible explanation is that these} particles are so large, and the downward flow so localized in the corners, that large particles are not capable of feeling these downward motions. {As a consequence PM particles larger than 10 mm mostly experience} upward motions, {hence remaining} near the top for long times until fluctuations {eventually} can get them down, or until they reach the central part of the tank where both mean flow and fluctuations vanish (figure \ref{fig2:coupeuz}f,h). This is why the contribution of the the central part to the downward flux of particles increases with increasing the particle diameter. 

This steric hypothesis was again tested numerically using the same 2D model as previously described by restricting the motion of the large particles in a sub volume $\vec X = (x,z) \in [-L_d/2,L_d/2] \times [0,H]$, assuming that $L_d=L-d_p$ in the case of real particles. It was observed that the mean height saturate at a higher value than $H/2$, which was not very surprising because concentrating particles at smaller values of $x$ is somewhat similar to imposing $\alpha<0$ in the previous case. However the increase observed in the numerics was at most about $20\%$, which is much smaller than the one observed experimentally. Whether this is due to the strong localization of downward motions for the present flow or if an interaction between the disc and the particles is responsible for it remains an open question although the fact that the effect is visible for PC18 particles is a good indication that the disc is not playing an important role in the problem.

\section{Conclusion}
\label{sec:concl}
We presented an experimental study of the dynamics of large heavy particles in a closed swirling flow, in a configuration for which gravitational effects are in competition with turbulent {agitation}. Particles with five diameters, two densities, were considered for a wide range of Reynolds numbers by varying the large scale forcing of the flow in the fully turbulent regime, so that the magnitude of the fluctuations remains proportional to the mean flow components. For all these parameters, particle trajectories were computed by particle tracking velocimetry so that their position PDFs and Eulerian mean flow field could be reconstructed. This allowed to explore the regime of strong confinement, for which the fluid velocity is at most of the order of the settling velocity of the particles, as well as the regime of loose confinement when the fluid velocity dominates over settling.

The strong confinement regime corresponds to a classical {suspension} regime for which the position PDFs are strongly linked to the mean flow topology whatever the size and density of the particles. In this regime for which gravity dominates over turbulent agitation and centrifugation, the mean height of the particles, $\overline{z}/H$, is proportional to the inverse of the turbulent P\'eclet number $Pe_\mathrm{turb} = v_s H /R^2 \Omega$. Although particles are much larger than any turbulent length scale, with response times of the order of the flow integral time-scale, such regime is well explained by the turbulent diffusion approach \cite{rouse1937,barresi1987,ayazi1989,magelli1990} which is usually valid only for very small particles.
 
For a given particle type, the loose confinement regime holds when {$Pe_\mathrm{turb} \leq 2$ in the present setup. In this regime, for which fluctuations dominate over gravitational effects, particles can explore the entire flow volume and very different behaviors are observed depending on the particle size. Particles smaller than a critical size, whose value lies between 6 and 10 mm, have a similar dynamics as the one observed in the strong confinement regime. However, inertial effects are more and more prominent when increasing the large scale velocity of the flow, at the origin of a quite strong centrifugation of the particles. The most important result is that particles larger than the critical size have a much stronger probability to remain {suspended} in the upper part of the vessel providing turbulent agitation allows them to reach the top of the tank so that the mean height can be $0.8~H$. Such preferential sampling is much stronger than the one observed in a similar setup with 2 discs rotating \cite{machicoane:njp2014} and it is of a different origin. Here particles are not found in torii corresponding to the neutral lines of the flow nor in the low fluctuation regions \cite{machicoane2016stochastic,machico_2021}. 

Investigation of the particle {density} current {suggests} that large particles sample preferentially upper regions due to geometrical constraints. Particles are too large to reach descending flow regions which are localized in the corners {of the vessel} in the present geometry. {As they mainly feel mean upward motions larger than their settling velocity, their altitude PDF becomes nearly exponential due to the presence of turbulent fluctuations, and are preferentially found near the top as if gravity was reversed. Very large particles then have erratic horizontal motions near the top until they {eventually} reach the central region where the mean flow vanishes, or until a turbulent fluctuation gets them down.} Finally, we note that as measurements were performed in the fully turbulent regime, it is impossible to disentangle between advection from the mean flow and from the fluctuations because the turbulence intensity is constant when varying the flow forcing. Future work could include a study with variable turbulence level by varying the fluid viscosity \cite{machico_2021}. Such study could help understanding how such large objects are {suspended} {when the mean flow dominates over the fluctuations}.

\begin{acknowledgments}
This work was supported by the French research programs ANR-16-CE30-0028, and IDEXLYON of the University of Lyon in the framework of the French program "Programme Investissements d'Avenir" (ANR-16-IDEX-0005).
\end{acknowledgments}

\bibliography{biblio_vkgb.bib}
	
\end{document}